\documentclass{article}

\usepackage{arxiv}

\usepackage[utf8]{inputenc} % allow utf-8 input
\usepackage[T1]{fontenc}    % use 8-bit T1 fonts
\usepackage{hyperref}       % hyperlinks
\usepackage{url}            % simple URL typesetting
\usepackage{booktabs}       % professional-quality tables
\usepackage{amsfonts}       % blackboard math symbols
\usepackage{nicefrac}       % compact symbols for 1/2, etc.
\usepackage{microtype}      % microtypography
\usepackage{lipsum}		% Can be removed after putting your text content
\usepackage{graphicx}
\usepackage[numbers,square,sort&compress]{natbib}
\usepackage{doi}
\usepackage{bm}

\usepackage{authblk}
\usepackage{amsmath}
\usepackage{algorithm}    % Include the algorithm package
\usepackage{algpseudocode}% Include the algorithmicx package's pseudocode stylings
\usepackage{todonotes}
\usepackage{float}
\usepackage{comment}

\title{THOI: An efficient and accessible library for computing higher-order interactions enhanced by batch-processing}

\author[1,2,6,*]{Laouen Belloli}
\author[3,4]{Pedro Mediano}
\author[5]{Rodrigo Cofré}
\author[1,2]{Diego Fernandez Slezak}
\author[6,*]{Rubén Herzog}

\affil[1]{\small Laboratorio de Inteligencia Artificial Aplicada, Instituto de Ciencias de la Computación, Universidad de Buenos Aires, Buenos Aires C1428EGA, Argentina}
\affil[2]{\small Departamento de Computación, FCEyN, UBA, Buenos Aires, Argentina}
\affil[3]{\small Department of Computing, Imperial College London}
\affil[4]{\small Division of Psychology and Language Sciences, University College London}
\affil[5]{\small Paris-Saclay University, CNRS, Paris-Saclay Institute of Neuroscience (NeuroPSI), 91400 Saclay, France}
\affil[6]{\small Institut du Cerveau, Paris Brain Institute, ICM, Inserm, CNRS, Sorbonne Université, 75013, Paris, France}
\affil[*]{\small ruben.herzog@icm-institute.org, laouen.belloli@gmail.com}

\date{}

\renewcommand{\headeright}{}
\renewcommand{\undertitle}{}
\renewcommand{\shorttitle}{\textit{THOI} Torch High-Order Interactions}

\hypersetup{
pdftitle={THOI: An efficient library for computing higher-order interactions enhanced by batch-processing},
pdfsubject={q-bio.NC, q-bio.QM},
pdfauthor={Laouen Belloli, Pedro Mediano, Rodrigo Cofré, Diego Slezak, Rubén Herzog},
pdfkeywords={Information theory, High-order interactions, O-information, Synergy, Redundancy, PyTorch},
}

\begin{document}
\maketitle

\begin{abstract}

% Propuesta Rodriigo 27/11/2024
Complex systems are characterized by nonlinear dynamics, multi-level interactions, and emergent collective behaviors. Traditional analyses that focus solely on pairwise interactions often oversimplify these systems, neglecting the higher-order interactions critical for understanding their full collective dynamics. Recent advances in multivariate information theory provide a principled framework for quantifying these higher-order interactions, capturing key properties such as redundancy, synergy, shared randomness, and collective constraints. However, two major challenges persist: accurately estimating joint entropies and addressing the combinatorial explosion of interacting terms. To overcome these challenges, we introduce THOI (Torch-based High-Order Interactions), a novel, accessible, and efficient Python library for computing high-order interactions in continuous-valued systems. THOI leverages the well-established Gaussian copula method for joint entropy estimation, combined with state-of-the-art batch and parallel processing techniques to optimize performance across CPU, GPU, and TPU environments. Our results demonstrate that THOI significantly outperforms existing tools in terms of speed and scalability. Specifically, THOI reduces the time required to exhaustively analyze all interactions in small systems ($\leq$ 30 variables). For larger systems, where exhaustive analysis is computationally impractical, THOI integrates optimization strategies that make higher-order interaction analysis feasible. We validate THOI’s accuracy using synthetic datasets with parametrically controlled interactions and further illustrate its utility by analyzing fMRI data from human subjects in wakeful resting states and under deep anesthesia. Finally, we analyzed over 900 real-world and synthetic datasets, establishing a comprehensive framework for applying higher-order interaction (HOI) analysis in complex systems. THOI opens new perspectives for testing both established and novel hypotheses about the multi-level, nonlinear, and multidimensional nature of complex systems.

\end{abstract}

\keywords{Information theory \and High-order interactions \and Synergy \and Redundancy \and PyTorch}

\section{Introduction}
\label{sec:introduction}
Complex systems are characterized by nonlinear dynamics, intricate interactions spanning multiple organizational levels, and emergent collective behaviors that cannot be fully explained by the properties of individual components \cite{jensen1998self, bar2019dynamics, boccara2010modeling, mitchell2006complex}. Understanding these systems is challenging due to the presence of complex, often poorly understood interdependencies, which traditional model-based approaches struggle to address.

Conventional methodologies typically focus on pairwise interactions, relying on simplifying assumptions that fail to capture the richness of these interdependencies. While pairwise models can provide useful insights, they neglect higher-order dependencies—interactions involving three or more elements simultaneously—that are critical for explaining the emergent phenomena observed in complex systems. By reducing the analysis to pairwise relationships, such approaches risk producing an incomplete or even misleading representation of the system's true dynamics and underlying processes \cite{battiston2022higher, battiston2020networks, boccaletti2023structure}.

To address the limitations of traditional pairwise approaches, information theory provides a powerful and general framework for quantifying the informational structure of complex systems. By extending Shannon's mutual information to account for higher-order interactions (HOI), this framework enables the identification and quantification of statistical dependencies that go beyond linear and pairwise correlations \cite{rosas2019quantifying, varley2023multivariate, timme2014synergy, james2017multivariate}. These higher-order dependencies are essential for understanding the emergent behaviors and intricate interdependencies inherent in complex systems.

At the heart of information theory lies the concept of entropy, which measures the unpredictability or randomness of a system and represents the average amount of information gained from observing it \cite{shannon1948mathematical, cover1991information}. In neuroscience, for instance, entropy has been applied to evaluate neural variability \cite{strong1998entropy, rieke1993coding} and to capture the complexity of information processing in the brain \cite{casali2013theoretically, tononi1994measure, castro2024dynamical}. This makes information theory particularly well-suited for studying the intricate dynamics of systems where higher-order dependencies play a critical role.

HOI refer to interdependencies involving three or more variables, capturing collective behaviors that cannot be reduced to pairwise relationships. By accounting for these interactions, HOI provide a deeper understanding of complex dependencies within systems. Multivariate information theory offers a rigorous framework to study HOI through several extensions of mutual information, including the total correlation ($TC$), dual total correlation ($DTC$), O-information ($\Omega$), and S-information \cite{rosas2019quantifying} (see Supplementary Methods). Each of these metrics reveals distinct aspects of higher-order dependencies: $TC$ quantifies collective constraints, $DTC$ represents shared randomness, $\Omega$ captures the balance between synergy and redundancy, and S-information reflects the overall level of interdependence.

Synergy and redundancy are key components of HOI. Synergy refers to information that emerges only when the system is analyzed as a whole and cannot be inferred from individual parts, while redundancy represents repeated information distributed across the system. These measures are derived from various linear combinations of low- and high-order entropies, unified under the entropy conjugation framework \cite{rosas2024characterising}.

In this work, we focus on $\Omega$, as it uniquely assesses the quality of interactions, i.e., synergy or redundancy dominance, rather than simply measuring the overall level of interdependence. Since its introduction, $\Omega$ has provided novel insights across diverse fields, including whole-brain dynamics \cite{gatica2021high, varley2023multivariate}, altered states of consciousness \cite{herzog2024high, kumar2024changes}, spiking neural networks \cite{stramaglia2021quantifying, newman2022revealing}, macroeconomic trends \cite{scagliarini2023gradients}, and music analysis \cite{scagliarini2022quantifying, medina2021hyperharmonic}.

Despite its broad applicability, the use of $\Omega$ faces two significant challenges: (i) like other information-theoretic measures, it requires the estimation of joint probability distributions, which often necessitates large datasets that may be difficult to obtain; and (ii) the combinatorial explosion of possible HOI, which scales exponentially with the number of variables (e.g., a system with 30 elements yields approximately $2^{30} \sim 10^9$ HOI).

To mitigate some of the challenges in analyzing complex systems, several open-source libraries have been developed for estimating entropy, mutual information, and related measures \cite{lizier2014jidt, ver2000non, ince2017statistical, HOI_toolbox, neri2024hoi}. While some of these tools include estimators for $\Omega$ and other HOI, they are often not optimized for large-scale analyses or for seamless accessibility in standard computational environments.

To address these limitations, we introduce THOI (Torch-based High-Order Interactions), a novel Python library specifically designed for efficient computation of HOI in large systems. THOI leverages the Gaussian copula (GC) method \cite{ince2017statistical, ma2011mutual} for joint entropy estimation, which enables direct computation from the covariance matrix of GC-transformed data. This approach bypasses the need for direct probability distribution estimation, significantly reducing computational complexity (see Supplementary Methods). To further enhance efficiency, THOI integrates with PyTorch \cite{paszke2019pytorch}, using optimized batch matrix operations to exploit the parallel processing capabilities of modern hardware, including CPUs, GPUs, and TPUs.

We evaluated THOI’s performance by comparing its computational efficiency to existing open-source libraries for estimating $\Omega$. For large systems where exhaustive computation of all possible interactions is infeasible, we implemented and validated optimization algorithms that balance accuracy and scalability using synthetic datasets with known ground-truth $\Omega$ values.

Finally, we demonstrate the practical utility of THOI through two applications: analyzing functional magnetic resonance imaging (fMRI) data to reveal reductions in synergistic interactions during deep anesthesia compared to wakefulness, and benchmarking its efficiency by analyzing over 900 multivariate datasets (both real-world and synthetic) in under 30 minutes on a standard laptop.

\section{Results}
\label{sec:result}
We validated THOI by benchmarking its computational efficiency and scalability against existing open-source libraries, focusing on exhaustive HOI computations. For larger systems where exhaustive analysis is infeasible, we developed and tested optimization strategies, demonstrating their effectiveness on both synthetic and real-world datasets. Additionally, we highlight THOI’s accessibility by analyzing over 900 distinct datasets in under 30 minutes on a standard laptop, showcasing its compatibility with commonly available computational setups.

\begin{figure}[h!]
    \centering
    \includegraphics[width=\textwidth]{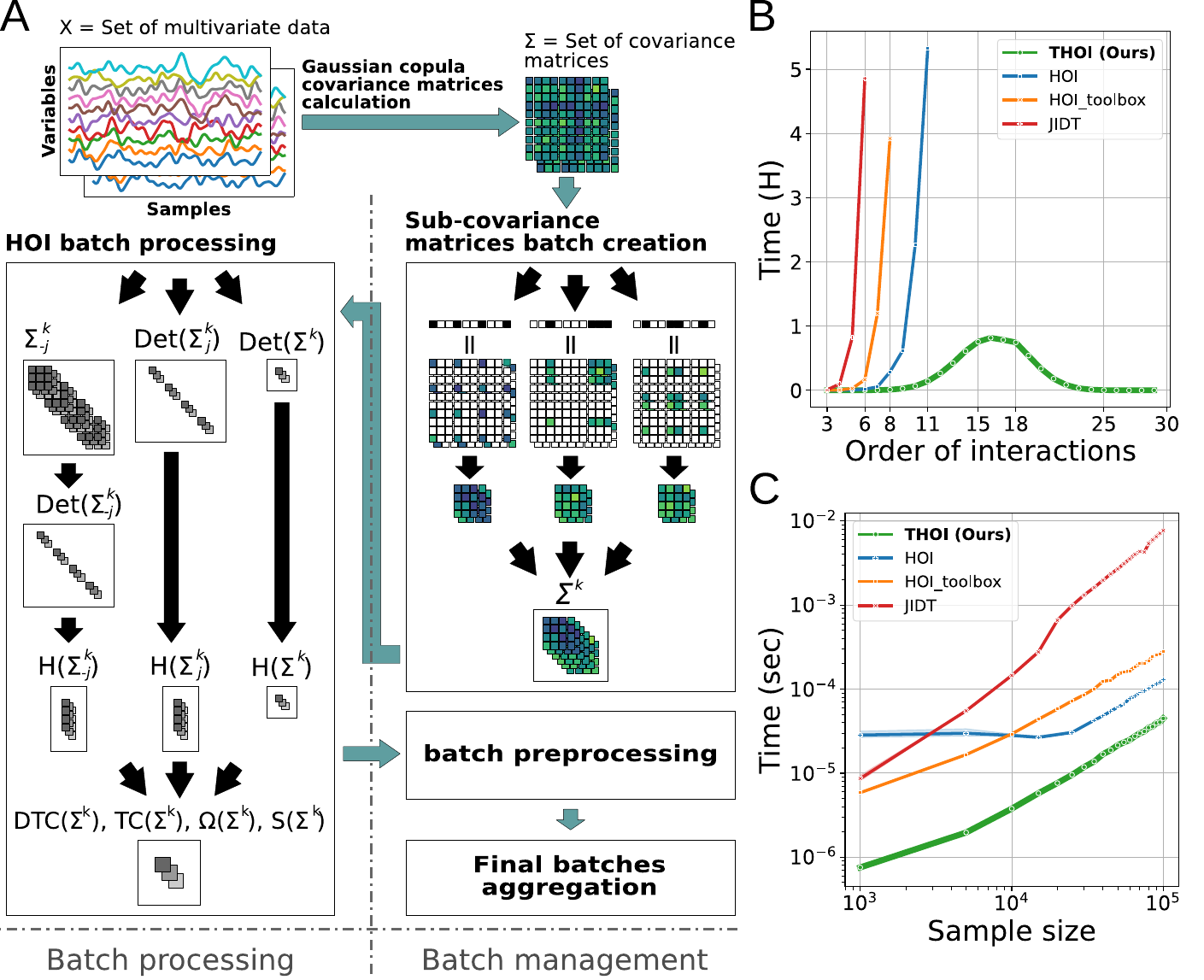}
    \caption{\small\textbf{Efficient computation of HOI using batch processing of covariance matrices.}  
    \textbf{A)} A set of multivariate time series $ X $ is transformed using the Gaussian copula approach, generating covariance matrices $\Sigma$ for each dataset. The covariance matrices are then sub-sampled using a batch of $ k $-plet indices, defined by a binary mask applied to $ \Sigma $, yielding the sub-covariance matrices $ \Sigma^k $ for each k-plet. These sub-covariance matrices are then batched together, and their determinants are computed, which are subsequently used to calculate the entropies and associated HOI defined by the $DTC$, $TC$, $ \Omega $, and $S$-information metrics, where the entropy is computed from the determinants of single variables $H(\Sigma^k_j)$, the whole system $H(\Sigma^k)$ and the whole system without a single variable  $H(\Sigma^k_{-j})$ (see  \ref{sup:generalization_of_MI} for detailed descriptions)  . Finally, batches are pre-processed using a custom function (e.g., extracting the minimum $\Omega$), and the results are aggregated to produce the final output. Note that multiple datasets with identical system and sample sizes can be processed simultaneously and the batch management system allows flexible analysis on the fly. 
    \textbf{B)} Computational time versus order of interactions for a 30-variable system with 1000 samples. The THOI method successfully computes all possible HOI in less than 6 hours, whereas other libraries are unable to process interactions beyond order 11 within the same time frame. 
    \textbf{C)} Log-log plot of computational time as a function of sample size for a 20-variable system. All libraries exhibit logarithmic scaling, but THOI outperforms the others in terms of computational speed.}
    \label{fig:batch_process}
\end{figure}

\subsection{Leveraging Gaussian entropy estimators with batch processing}

To address the combinatorial explosion in computational complexity as the number of variables increases (see Supplementary Material, Section \ref{sup:combinatorial_explosion}), we implemented a PyTorch-based batch-processing architecture \cite{paszke2019pytorch}. This approach groups and processes data in parallel, significantly improving efficiency in the computation of HOI across large datasets (see Methods \ref{sec:efficient_computation}).

The workflow (Figure \ref{fig:batch_process}A) begins by transforming multivariate time series $X$ into covariance matrices $\Sigma$ using the Gaussian copula method (see Supplementary Methods, Section \ref{sup:gaussian_copula}). Binary masks are then applied to extract sub-covariance matrices $\Sigma^k$ for each $n$-plet of $k$ variables. These matrices are processed in batches to compute the determinants required for estimating entropies and HOI metrics such as $DTC$, $TC$, $\Omega$, and $S$-information (see Section \ref{sup:generalization_of_MI}).

This architecture enables the simultaneous processing of multiple datasets, allowing for real-time analyses such as identifying the $n$-plet that minimizes the average $\Omega$ across datasets. Additionally, users can easily control the batch size via a single parameter (\textit{batch size}), ensuring scalability, efficient memory management, and compatibility with standard computational platforms.

A key feature of THOI is its ability to handle sub-covariance matrices of varying sizes within the same batch. Since different orders of interactions correspond to matrices of different dimensions, traditional batch processing becomes inefficient due to the fixed size of each batch, requiring iterative loops for matrices of different sizes. To overcome this limitation, we implemented independent variable padding, allowing computations for different interaction orders to be performed efficiently within the same batch (see Methods \ref{sec:padded_batches}). 

\subsection{Enhanced Performance}

To evaluate the computational advancements of THOI, we compared its performance against three widely used open-source libraries for computing HOI:

\paragraph{HOI\_toolbox (Higher-Order Interactions Toolbox)} \cite{HOI_toolbox}: A Python library optimized for efficiently computing HOI in datasets using the Gaussian copula estimator \cite{ince2017statistical}. It is designed for memory and processing efficiency but is limited to systems with $N \leq 20$ for exhaustive computations.

\paragraph{HOI (High-Performance Estimation of Higher-Order Interactions)} \cite{neri2024hoi}: A Python library designed to leverage high-performance computing (HPC) architectures. While it supports multiple estimators, it does not specifically address the combinatorial challenges associated with large-scale HOI computations.

\paragraph{JIDT (Java Information Dynamics Toolkit)} \cite{lizier2014jidt}: A Java-based library primarily focused on analyzing information flow dynamics. It includes the Kraskov-Stögbauer-Grassberger (KSG) estimator \cite{kraskov2004estimating} for entropy and related measures but does not tackle the combinatorial explosion problem inherent to HOI.

For benchmarking, we created a multivariate Gaussian system with 30 variables, each independently drawn from $\mathcal{N}(0,1)$ (it is worth noticing time performance is not affected by the variable distribution and we only report the distribution for completeness). The goal was to compute $\Omega$ for all combinations of variables from order 3 to 30 without leveraging additional parallelization, ensuring a fair comparison. All computations were performed on a standard laptop equipped with an Intel Core i9 processor, 64 GB of RAM, running Linux Mint 21.

We observed a clear performance improvement with THOI compared to the other libraries (Figure \ref{fig:batch_process}B). Using THOI, we were able to compute all interaction orders from 3 to 30 in just 5.8 hours. In contrast, the other libraries were limited to order 8 within a reasonable computational time (<5 hours). The maximum computational time was expected to be close to 15, as the combinatorial explosion occurs at approximately $N/2$.

While computing all possible interactions with the other libraries was infeasible due to time or memory constraints, we extrapolated their computational times. For \textbf{HOI\_toolbox}, this would have taken approximately 221 days. \textbf{HOI} would have required 2 days, with a memory overload of 240 GB just for $n$-plets indices creation and allocation. \textbf{JIDT} would have taken roughly 17 years to complete. Additionally, THOI can compute $TC$, $DTC$, $S$-information, and $\Omega$ in a single run, whereas other libraries require separate functions for each measure, which can increase computational time by a factor of 4. In terms of memory usage, THOI completed the full set of interaction orders using less than 3 GB of memory (with a batch size of 10,000). Finally, we assessed how computational time scales with sample size in a system of 20 variables. Again, THOI outperformed the other libraries in terms of computational efficiency (Figure \ref{fig:batch_process}C).

\subsection{Optimization for larger systems: greedy and simulated annealing algorithms}

Although THOI greatly accelerates the computation of HOI, exhaustively assessing all possible combinations in larger systems is computationally prohibitive. To address this, heuristic optimization algorithms, such as greedy algorithms (GA) \cite{herzog2022genuine} and simulated annealing (SA) \cite{varley2023multivariate, hourican2024efficient}, have been implemented in the THOI framework. These algorithms optimize an objective function --here  $\Omega$ for demonstrative purposes-- by targeting the most relevant combinations of variables, circumventing the need for exhaustive enumeration. 
The GA, deterministic in nature, optimizes locally at each order of interaction, using its outcome as the starting point for the subsequent order of interaction. In contrast, SA leverages stochasticity to escape local optima by allowing random changes early in the search, gradually refining solutions by reducing randomness over time. Our implementation of the GA optimizes across multiple initial conditions, while the SA method employs both within-order optimization (focusing on a specific order of interaction) and across-order optimization (allowing transitions between solutions of different order of interaction).

\begin{figure}[h!] % Positioning options: here, top, bottom, page (htbp)
    \centering
    \includegraphics[width=\textwidth]{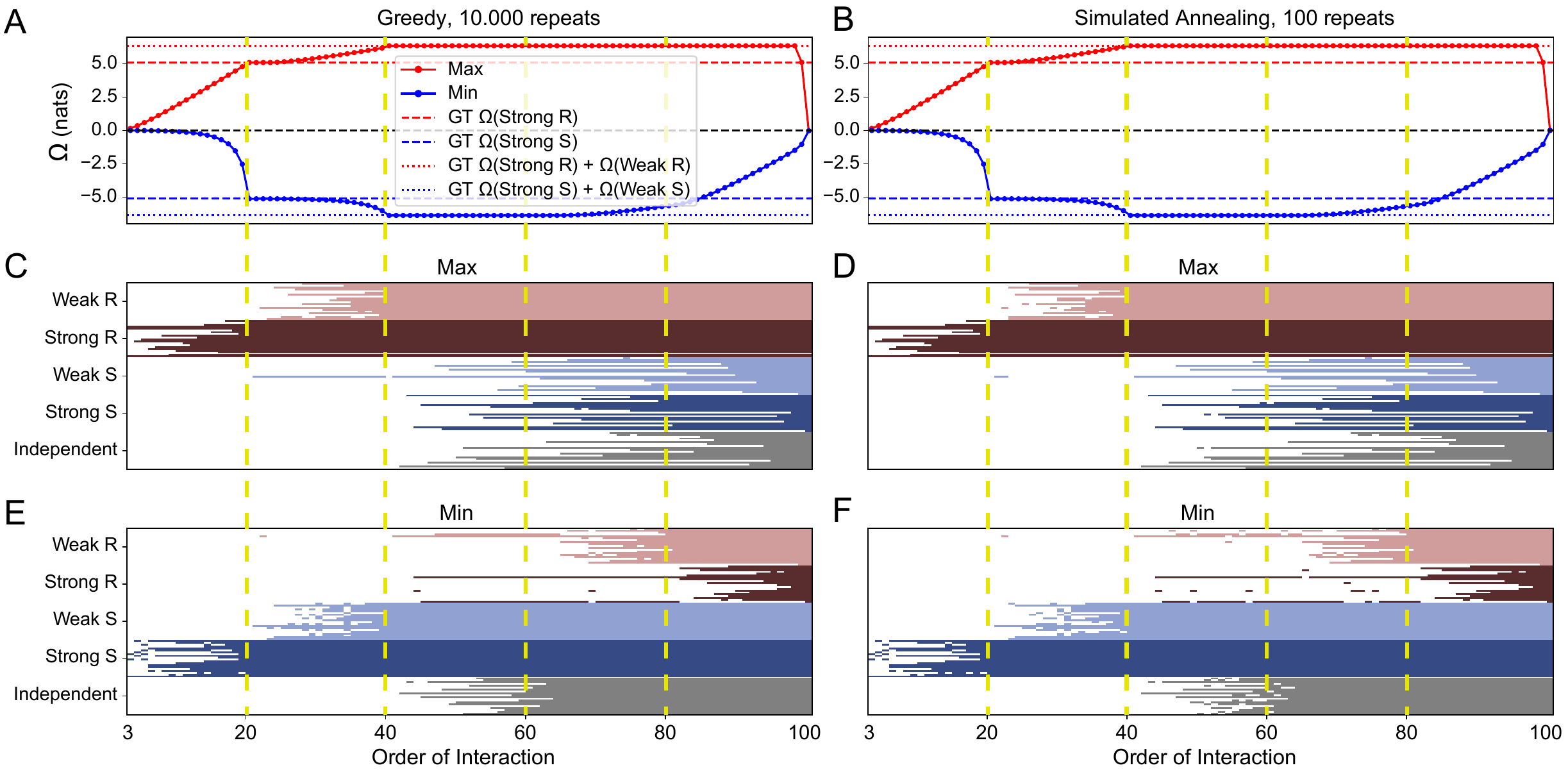}
    \caption{\small\textbf{Within-order optimization with greedy and simulated annealing algorithms} 
    \textbf{A, B)} Maximum (red) and minimum (blue) $\Omega$ obtained by greedy and SA algorithms for a $100$-variable system composed of strong/weak R and S systems and an independent system, each with 20 variables. Dashed horizontal lines indicate ground truth for the weak systems, and dotted lines represent the sum of weak and strong systems (red for R, blue for S). Both algorithms successfully identify the systems, but greedy required $100$ times more repeats than SA.
    \textbf{C, D)} Subsets of variables that maximize $\Omega$ at each order of interaction for greedy and SA. Systems are color-coded (weak R: light red; strong R: dark red; weak S: light blue; strong S: dark blue; independent: gray). Both algorithms prioritize strong R, then weak R, followed by a mix of independent and S systems.
    \textbf{E, F)} Subsets of variables that minimize $\Omega$. Both algorithms prioritize strong S, then weak S, followed by a combination of independent and R systems, with a preference for the former.}
    \label{fig:fig6_greedy_and_SA_vs_N}
\end{figure}

To evaluate the accuracy of these algorithms, we designed a system of $100$ variables consisting of five $20$-variable sub-systems where the ground truth is known \cite{rosas2024characterising} (see Supplementary Methods \ref{sup:PMGs}): two redundant (R) systems (weak and strong), two synergistic (S) systems (weak and strong), and one independent system. The strength of each sub-system was determined by the coupling parameter $c$, weak = $0.5$ and strong = $1$, (see Supplementary Methods \ref{sup:PMGs}). Following the additive property of the $\Omega$ for independent components (i.e. $A \perp B \Rightarrow \Omega(A+B) = \Omega(A) + \Omega(B)$), this configuration allowed us to pre-define the maximal and minimal $\Omega$ values, corresponding to the sum of R and S sub-systems, respectively, and to ensure perfect synergy-redundancy balance ($\Omega$=0) for the whole 100-variable system.

We ran the GA and the SA for each order of interaction, finding the correct identity (i.e. the variables of the $n$-plets) and $\Omega$ values associated with each of the R and S subsystems with both algorithms (Figure \ref{fig:fig6_greedy_and_SA_vs_N}).
% Although the greedy approach required 10000 different initial conditions to achieve the same solution that the SA achieved with just 100 repeats, the SA required almost 2 orders of magnitude more time for the same number of repeats (Supplementary Figure XX).
% This was due to the fact that the SA algorithm optimizes for each order separately, while the greedy algorithm uses the solution of the order of interaction $k$ as initial condition for the order of interaction $k+1$.
As we progressed from lower to higher orders, the maximum $\Omega$ increased monotonically, reaching order $20$ (the size of the strong R system) and then slowed in its rise before saturating at order $40$ (strong R + weak R). After this saturation point, the first decline occurred at order $99$, when the entire weak S system was included, and then at order $100$, when both S systems were completely included, achieving a perfect balance between synergy and redundancy (Figure \ref{fig:fig6_greedy_and_SA_vs_N}C, D) expected by design. 
Conversely, minimum $\Omega$ increased non-linearly up to $20$ (strong S system), then at a slower rate until saturating at $40$ (strong S + weak S). Beyond this point, the minimum $\Omega$ started to increase at order $60$, where variables from the weak R system were included, continuing to decline as strong R variables were added, reaching zero (Figure \ref{fig:fig6_greedy_and_SA_vs_N}E, F).

Additionally, the greedy algorithm, with its bottom-up approach, required $20$ iterations to identify the correct maximum $\Omega$ solutions, whereas it needed two orders of magnitude more for the minimum, underscoring the higher-order and non-linear nature of synergies compared to redundancy (Supplementary Figure \ref{supp_fig:greedy_vs_repeats_N-30}).

\subsection{Analysis of human brain activity under anesthesia}
To demonstrate the utility of THOI in empirical research, we applied it to fMRI data from $16$ subjects, each undergoing both wakeful rest and deep anesthesia. The data, previously published in Ref.~\cite{nemirovsky2023implementation}, include brain activity across $55$ brain regions, which are organized into $11$ distinct brain networks, each containing five regions. This analysis was inspired by prior studies that have linked alterations in levels of consciousness to changes in brain complexity \cite{tononi1994measure, demertzi2019human, sarasso2021consciousness, nemirovsky2023implementation, castro2024dynamical}. Our primary goal was to investigate whether anesthesia induces significant changes in the $\Omega$, reflecting shifts in the brain's HOI during these distinct states of consciousness.

\begin{figure}[h!] % Positioning options: here, top, bottom, page (htbp)
    \centering
    \includegraphics[width=\textwidth]{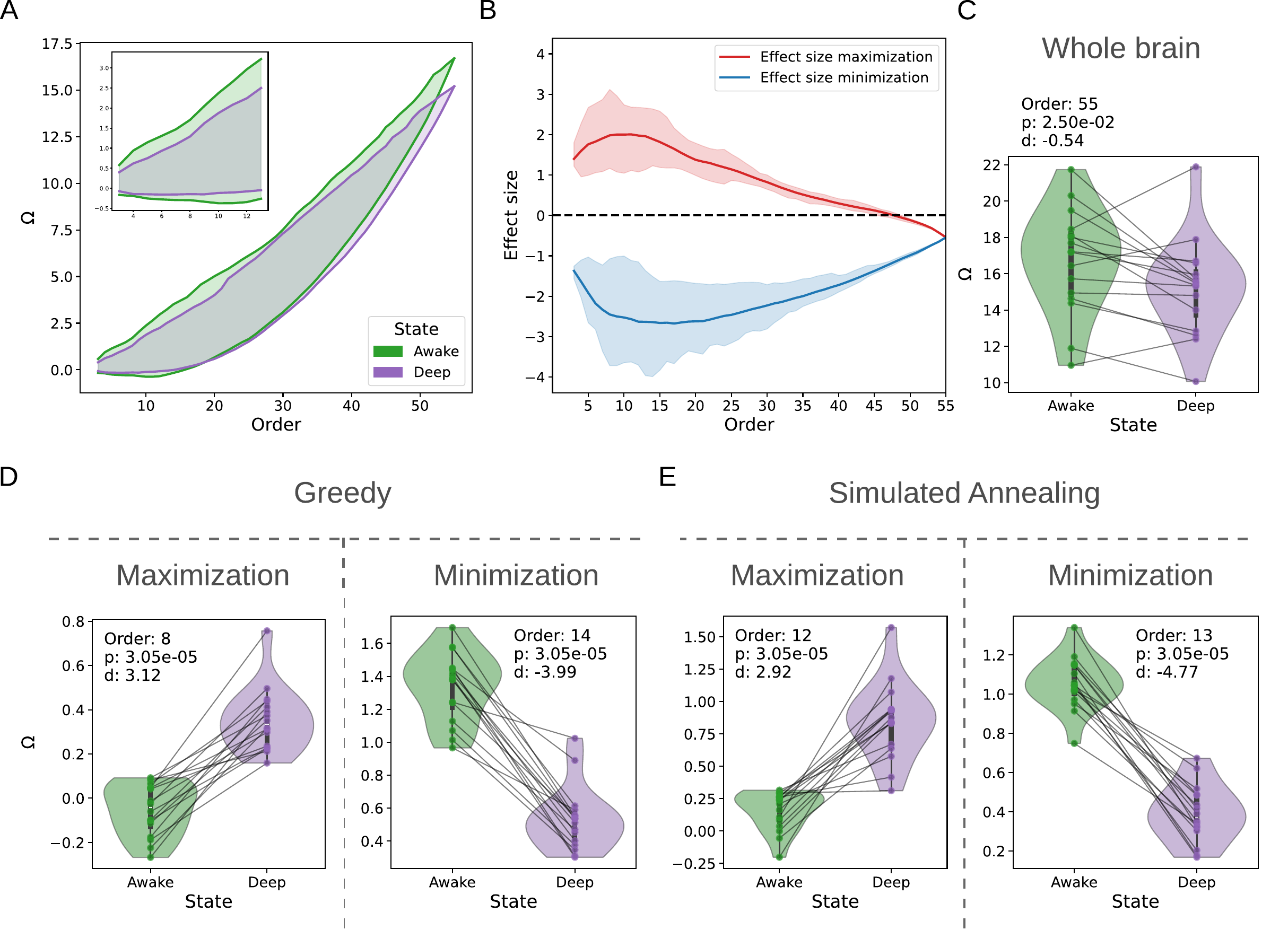}
    \caption{\small\small
    \textbf{A)} Estimated maximum and minimum $\Omega$ via the GA for awake (green) and deep anesthesia state (purple). Inset shows the reduction of minimum $\Omega$ at lower orders of interaction. 
    \textbf{B)} Average maximum (red) and minimum (blue) effect size obtained from a GA tailored to amplify the difference between the two conditions. Shaded areas denote the range from the minimum to the maximum value for each optimization procedure.
    \textbf{C)} Distribution of $\Omega$ for the whole-brain in awake (green) and deep anesthesia state (purple). Each dot is a subject and lines connect their respective value in both conditions. Despite the trend to reduce redundancy, no significant difference was found (Wilcoxon $p>0.001$)
    \textbf{D)} Distribution of the $n$-plets that maximizes (left) and minimizes (right) the effect size obtained by the GA. 
    \textbf{E)} Same as \textbf{D}, but for the $n$-plets obtained with the SA algorithm. Order is the number of elements in $n$-plets, $p$ is the Wilcoxon p-value and $d$ is the Cohen's $d$.   
    }
    \label{fig:fig8_anesthesia}
\end{figure}

We first applied the GA to independently obtain maximum and minimum $\Omega$ across different orders of interaction in both conditions (Figure \ref{fig:fig8_anesthesia}A). The results revealed that deep anesthesia led to a reduction in both the maximum and minimum $\Omega$ values across all orders of interaction, indicating a significant decrease in brain complexity during this state.
To identify groups of brain regions with the most pronounced differences between wakefulness and deep anesthesia, we employed a GA to maximize and minimize effect sizes (separately) in $\Omega$ across both conditions (Figure \ref{fig:fig8_anesthesia}B). While no significant differences were observed at the whole-brain level (Wilcoxon $p>0.001$), Figure \ref{fig:fig8_anesthesia}C), significant differences with large effect sizes were found (Cohen's $|d| > 3$, Wilcoxon test $p < 0.001$, not corrected) for both increases and decreases in $\Omega$ (Figure \ref{fig:fig8_anesthesia}D,E).

The most notable increase in $\Omega$ involved eight regions from distinct resting-state networks \cite{nemirovsky2023implementation}. This interaction, which was synergy-dominated during wakefulness, shifted to redundancy-dominated during anesthesia. Conversely, the largest decrease in $\Omega$ encompassed $14$ regions, including four from the Cingulo-Opercular network, four from the Cingulo-Parietal network, and two from the Default Mode network, among others. Despite the reduction in $\Omega$, these regions remained redundancy-dominated under both conditions. Using a SA algorithm across orders, we found results that mirrored those from the GA (Figure \ref{fig:fig8_anesthesia}E), showing significant differences with large effect sizes for both $\Omega$ increases and decreases. Although the $\Omega$ increase was less than with the GA, it similarly resulted in synergy-to-redundancy shifts in some subjects. Reductions in $\Omega$, again, did not result in synergy dominance, even with greater effect sizes than those found with the GA.

In summary, these findings suggest that anesthesia compresses the range of $\Omega$ values in the brain, reducing both maxima and minima, and diminishes the dominance of synergy across interactions between distinct networks while reducing redundancy within same-network regions. 
%This supports the hypothesis that maintaining conscious wakefulness requires complex integration across functionally specialized brain networks \cite{casali2013theoretically}.

\subsection{Analysis of large database of synthetic and real-world systems}
To demonstrate the potential of THOI and its application to a wide spectrum of complex systems, we analyzed 920 datasets, including both synthetic and real-world data, with system sizes ranging from 5 to 20 \cite{cliff2023unifying}. We exhaustively computed $TC$, $DTC$, $\Omega$, and $S$-information for all variable combinations (from 2 to $N$, the system size) across the entire database in under 20 minutes using a laptop with an Intel i7-13800H processor.
\begin{figure}[h!] % Positioning options: here, top, bottom, page (htbp)
    \centering
    \includegraphics[width=\textwidth]{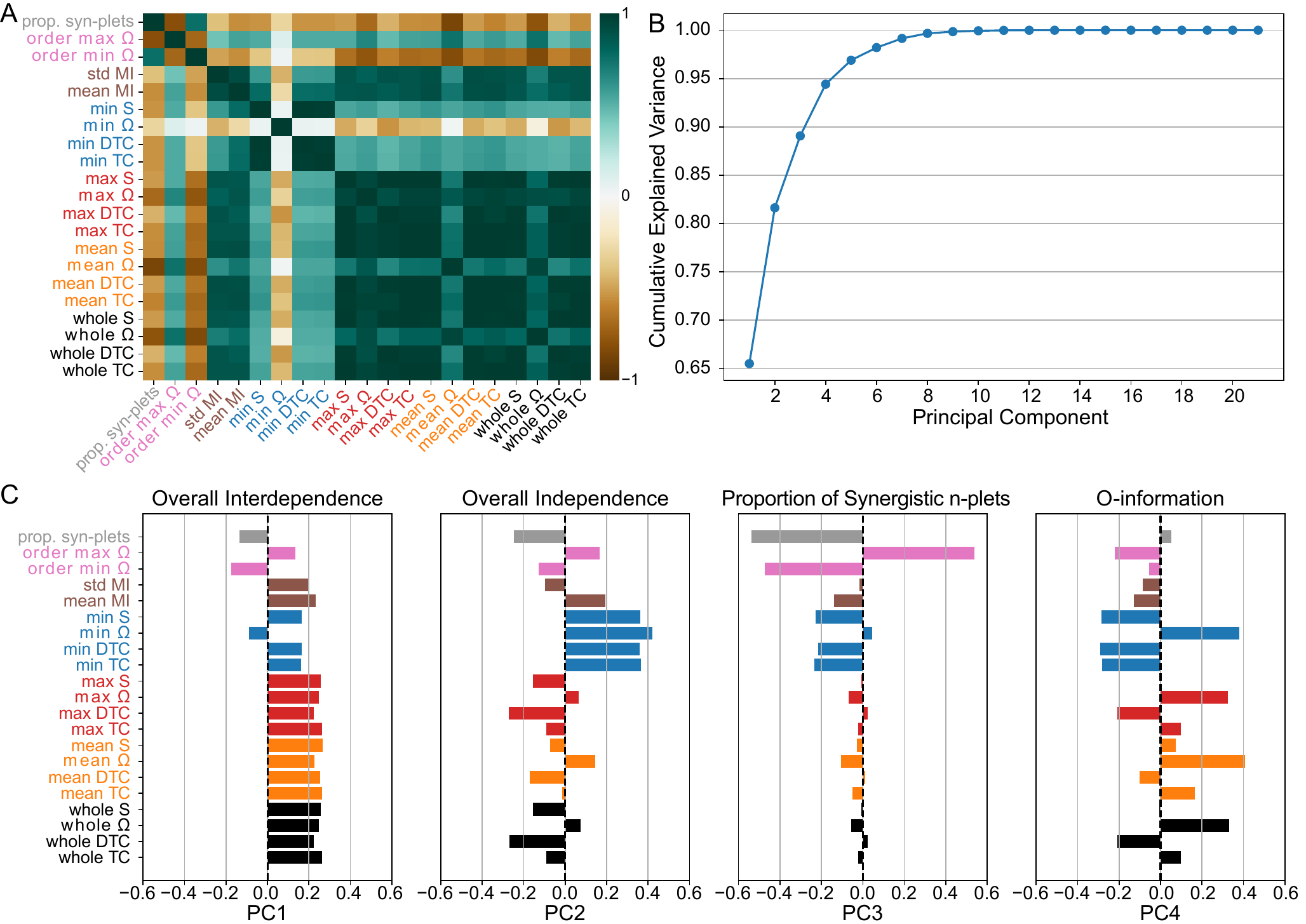}
    \caption{\small
    \textbf{A)} Spearman correlation matrix of features across datasets. Colors code different types of features. 'prop. sy$n$-plets' is the proportion of synergy-dominated $n$-plets out of the total number of $n$-plets. 'order max $\Omega$' and 'order min $\Omega$' is the order where  $\Omega$ was maximized and minimized, respectively, normalized by the system size. 'mean MI' and 'std MI' are the mean and standard deviation of the pairwise mutual information for each dataset. The prefix 'whole' indicates that the whole system was considered (i.e. all the system variables).
    \textbf{B)} Cumulative explained variance associated with each PC after PCA. The first four components capture approximately 95\% of the variance. 
    \textbf{C)} Values of the first four PCs on each feature. Colors are the same as in   
    \textbf{A}. PC1 captures the overall interdependencies, by grouping together all the 'max', 'mean', 'whole' and 'MI' related features. PC2 captures the overall independence by grouping together all the 'min' related features. PC3 captures the proportion of synergy-dominated interaction by grouping together 'prop. syn-plets' and the order at which $\Omega$ is maximized and minimized. PC4 captures the behavior of $\Omega$, by grouping together its maximum, minimum, mean and whole-system values.
    }
    \label{fig:hoi_features}
\end{figure}

For each dataset, we extracted features characterizing the informational structure, including the maximum, minimum, mean (across all orders of interaction) and whole-system values of the four aforementioned metrics, the mean and standard deviation of pairwise mutual information (MI), the order at which $\Omega$ is maximized and minimized (normalized by $N$), and the proportion of synergy-dominated $n$-plets relative to the total $n$-plets. Although these features were derived from an exhaustive computation of all interactions, all except the proportion of synergy-dominated $n$-plets can be efficiently estimated using GA or SA, making them applicable to larger systems. 

We found strong correlations among several features (Figure \ref{fig:hoi_features}A), particularly those related to the maximum, mean, whole-system, and pairwise mutual information (MI).
Using principal component analysis (PCA), we observed that the first principal component (PC) explained over 65\% of the variance (Figure \ref{fig:hoi_features}B), with values primarily reflecting features tied to the maximum, mean, whole-system, and MI (Figure \ref{fig:hoi_features}C). We termed this component 'Overall Interdependence' as increases in the grouped features indicate stronger overall interdependencies. The second PC primarily loaded onto features associated with the minimum, leading us to term it 'Overall Independence' as increases in these features reflect departure from independence. Importantly, the orthogonality of the PCs ensures that increases in overall interdependence do not imply decreases in overall independence, as these properties can vary within and across orders of interactions. The third PC primary loaded on features related to the proportion of synergy-dominated $n$-plets and to the order at which the $\Omega$ is maximized and minimized. We termed this PC 'Proportion of Synergistic $n$-plets', as measuring any of these 3 features will capture the proportion of synergy-dominated $n$-plets. It is remarkable that even when the proportion of synergistic $n$-plets can't be directly measured in large systems, our results suggest that it can be approximated by the order of the maximum or minimum $\Omega$.
Finally, the fourth PC (which in addition to the other three PCs accounts for approximately 95\% of the variance), loaded mainly into the maximum, minimum, mean and whole-system $\Omega$-related measures. Accordingly, we termed it 'O-information'.

These results highlight that THOI provides an efficient and ready-to-apply approach for estimating features that characterize HOI in complex systems. Our analysis suggests that these features capture key properties such as the overall interdependence, overall independence, the proportion of synergy-dominated interactions, and the synergy-redundancy balance. Together, these dimensions provide a general framework for understanding the informational structure of complex systems.

\section{Methods}
\label{sec:methods}

\label{sec:efficient_computation}
\subsection{Multivariate information theory}
\label{sup:Multivariate_information_theory}

We first introduce the fundamental concepts of information theory, focusing on entropy and mutual information, before extending these ideas to HOI, with particular emphasis on the $\Omega$. As previously noted, entropy serves as a cornerstone of information theory, quantifying uncertainty or information content.

In what follows, we denote by $ X^n $ a continuous multivariate random variable with $ n $ components (or simply $ X $ when referring to a single variable), and by $ X_j $ the $ j $-th component of $ X^n $. A specific realization of $ X $ is represented as $ x $, with $ p(x) $ denoting its associated probability. Given that we are working with continuous data, we utilize Shannon's differential entropy, denoted as $ H(X) $, defined as follows:

\begin{equation}
\label{eq:differential_entropy}
    H(X) = - \int_x p(x)\log(p(x)) dx
\end{equation}

This formula requires knowledge of $ p(x) $, i.e., an estimation of the probability density function (PDF). Unlike discrete entropy, differential entropy is unbounded and can take both negative and positive values, with units in nats when the natural logarithm is used. 

For simplicity, we will refer to differential entropy simply as entropy throughout this work. The equation \ref{eq:differential_entropy} is valid for multivariate systems, where $ H(X^n) $ denotes the joint differential entropy. This, in turn, depends on the joint probability density function (JPDF), and the integrals are taken over the entire support of the JPDF.

For a pair of continuous random variables $X$ and $Y$ (or a bivariate system $X^2$), the mutual information $I(X;Y)$ follows:

\begin{equation}
\label{eq:MI-tc}
    I(X;Y) = H(X) + H(Y) - H(X,Y) = \sum_{j=1}^2 H(X_j) - H(X^2)
\end{equation}

It can also be expressed in terms of conditional entropies:

\begin{equation}
\label{eq:MI-dtc}
    I(X;Y) = H(X,Y) - H(Y|X) - H(X|Y) = H(X^2) - \sum_{j=1}^2 H(X_j | X^2_{-j})  
\end{equation}

Here, $X^2_{-j}$ corresponds to the full system without the variable $X_j$. As previously mentioned, it informs about the uncertainty that is reduced on one variable when we know the other and is guaranteed to be non-negative. In the following, we present its generalizations for multivariate systems with 
$n>2$, where \ref{eq:MI-tc} and \ref{eq:MI-dtc} are no longer equivalent. 

\subsection{Generalizations of the mutual information}
\label{sup:generalization_of_MI}
The generalizations of \ref{eq:MI-tc} and \ref{eq:MI-dtc} for higher order interactions have been called the total correlation ($TC$) \cite{te1978nonnegative}, and the dual total correlation ($DTC$) \cite{watanabe1960information}. They follow: 

\begin{equation}
\label{eq:TC}
    TC(X^n) = \sum_{j=1}^n H(X_j) - H(X^n)
\end{equation}
\begin{equation}
\label{eq:DTC}
    DTC(X^n) = H(X^n) - \sum_{j=1}^n H(X_j|X^n_{-j})
\end{equation}

Both the $TC$ and the $DTC$ are multivariate generalizations of the mutual information \ref{eq:MI-tc} and \ref{eq:MI-dtc}, respectively.
However, they are not equivalent. They are non-negative quantities. $TC$ has been interpreted as collective constraints, while $DTC$ represents shared randomness.

The $\Omega$-information as proposed in \cite{rosas2019quantifying} quantifies high-order interdependencies in complex systems by measuring the balance between redundancy and synergy. 
The $\Omega$-information is defined as the difference between the total correlation and the dual total correlation as follows:

\begin{equation}
\label{eq:O-info}
    \Omega(X^n) = TC(X^n) - DTC(X^n) = (n - 2) H(X^n) + \sum_{j=1}^n [H(X_j) - H(X^n_{-j})]
\end{equation}

Therefore, $ \Omega $ can be expressed solely in terms of entropies. A system is said to be synergy-dominated if $ \Omega < 0 $, and redundancy-dominated if $ \Omega > 0 $. In other words, if $ DTC(X^n) $ exceeds $ TC(X^n) $, the system is considered more synergistic than redundant. This indicates that the interdependencies between the variables contribute information that cannot be inferred from examining the variables individually. Conversely, if $ TC(X^n) $ exceeds the $ DTC(X^n) $, the system is classified as more redundant than synergistic. This suggests that the interdependencies are largely due to shared or repeated information across the variables.

Finally, $S$-information is defined as the addition between the total correlation and dual total correlation as follows:

\begin{equation}
\label{eq:S-info}
    S(X^n) = TC(X^n) + DTC(X^n)
\end{equation}

\subsection{Efficient computation of HOI via Gaussian copulas}

The computational cost of calculating HOI is primarily driven by two factors: the need to compute the joint entropy for combinations of $k$ elements and the inherent combinatorial complexity of the problem (see Supplementary Methods \ref{sup:combinatorial_explosion}). Calculating $\Omega$ is computationally intensive, with efficiency varying depending on the entropy estimator used.

% For instance, the KSG estimators\cite{kraskov2004estimating} employs a non-parametric approach to approximate mutual information and its generalizations. It requires identifying the k-th nearest neighbor for each sample in a combination, with these neighbors dynamically changing depending on the subset of elements involved. Even when applying techniques where joint entropies are precomputed and reused to estimate $\Omega$—the computational cost becomes prohibitive for systems of moderate size.

In contrast, the entropy estimator for Gaussian variables, explained in Supplemental Methods \ref{sup:gaussian_entropy_formula}, simplifies this process by requiring the calculation of the covariance matrix only once for the entire system. Subsequent sub matrices associated with each combination of variables ($n$-plet) can then be extracted efficiently, making this approach computationally less expensive. However, the number of possible combinations and the associated computational cost scale exponentially, making this an NP-hard problem. Although it cannot be fully solved, certain computational strategies, such as parallelization and batch processing, can significantly improve both time and memory performance.

\subsubsection{Parallel and batched computation of HOI}
In this section, we present how the Gaussian copula entropy estimator can be leveraged to compute the measures defined in equations \ref{eq:TC}, \ref{eq:DTC}, \ref{eq:O-info}, and \ref{eq:S-info} for multiple subsets of variables ($n$-plets) and for multiple datasets of the same size in parallel and in a batched fashion using PyTorch. This implementation allows us to exploit the power of matrix operations and batch processing on both CPUs and GPUs, as well as the parallelization capabilities of HPC architectures. The provided Python library depends only on PyTorch and NumPy and works seamlessly with or without CUDA (the library for performing computations on GPUs), providing an open-source, accessible, and ready-to-use Python library similar to NumPy.

Batch processing is feasible due to the following reasons:

\begin{enumerate}
    \item \textbf{Independence of subsets}: The computations for different subsets of variables are independent, allowing for concurrent processing without interference.

    \item \textbf{Linear algebra operations}: The core computations involve linear algebra operations (e.g., matrix slicing, determinant calculation) that are inherently parallelizable and can be vectorized using tensor operations.

    \item \textbf{Advanced computing architectures}: Modern computing architectures available in ordinary laptops, such as GPUs and multi-core CPUs, are optimized for parallel computations on tensors, making batch processing computationally efficient.
\end{enumerate}

To understand the core computations and why they can be processed in parallel, we will decompose the algorithm shown in Figure~\ref{fig:batch_process} into its components. From equations \ref{eq:TC}, \ref{eq:DTC}, \ref{eq:O-info}, and \ref{eq:S-info}, we need to compute the following terms:

\begin{enumerate}
    \item $H(X_j)$: The entropy of each random variable $X_j$.
    \item $H(X^n)$: The entropy of the joint random variable $X^n$, i.e., the entire system.
    \item $H(X^n_{-j}) = H(X_1,\ldots,X_{j-1},X_{j+1},\ldots,X_n)$: The entropy of the entire system excluding $X_j$.
\end{enumerate}

Since all these terms are entropies of the form given in equation~\ref{eq:differential_entropy}, we can use the Gaussian copula approach (equation~\ref{eq:gaussian_copula_covariance_estimator} in Supplemental Methods) to estimate the full system's covariance matrix and extract the required $n$-plet sub-covariance matrices as a batch. Once we have the sub-covariance matrices, we can compute Gaussian entropies using equation~\ref{eq:gaussian_differential_entropy} in a batched fashion, as all the operations are matrix computations available for batch processing in PyTorch. Finally, we compute the measures from equations \ref{eq:TC}, 
 ~\ref{eq:DTC}, \ref{eq:O-info}, and \ref{eq:S-info} using only addition and subtraction operations.

To construct batches of sub-covariance matrices from the $n$-plets, we proceed as follows:

\begin{enumerate}
    \item \textbf{Covariance matrices}: We consider a set of $D$ covariance matrices $\mathbf{\Sigma}_d \in \mathbb{R}^{D \times N \times N}$ represents the covariance matrix of a dataset or a random vector of dimension $N$.

    \item \textbf{$n$-plets}: We define a collection of $B$ $n$-plets $\{ \mathbf{i}_b \}_{b=1}^B$, where each $\mathbf{i}_b \in \mathbb{Z}^K$ is an index set representing a subset of $K$ variables out of $N$.

    \item \textbf{Batch formation}: We create a batch by pairing each $n$-plet with each covariance matrix, resulting in $B \times D$ combinations. This pairing is represented using high-dimensional tensors, enabling simultaneous processing.

    \item \textbf{Sub-covariance extraction}: For each combination, we extract the sub-covariance matrix corresponding to the variables in the $n$-plet from the full covariance matrix. This extraction is achieved through tensor indexing operations, ensuring efficiency and parallelizability
\end{enumerate}

The following section describes in more detail the batch operations required to create the batched sub-covariance matrices.

\subsection{Batched sub-covariance matrices}
\label{sec:batched_subcovariance_matrices}

Let $\mathbf{\Sigma} \in \mathbb{R}^{D \times N \times N}$ be the tensor of covariance matrices and $\mathbf{I} \in \mathbb{Z}^{B \times K}$ be the tensor of $n$-plets.

\begin{enumerate}
    \item \textbf{Tensor expansion}:

    \begin{itemize}
        \item \textbf{Expand $n$-plet Indices}: We expand the $n$-plet indices to align with the dimensions of the covariance matrices:
        \[
        \mathbf{I}_{\text{exp}} \in \mathbb{Z}^{B \times D \times K},
        \]
        where $\mathbf{I}_{\text{exp}}[b, d, :] = \mathbf{i}_b$ for all $d$.

        \item \textbf{Expand covariance matrices}: We expand the covariance matrices to align with the batch of $n$-plets:
        \[
        \mathbf{\Sigma}_{\text{exp}} \in \mathbb{R}^{B \times D \times N \times N},
        \]
        where $\mathbf{\Sigma}_{\text{exp}}[b, d, :, :] = \mathbf{\Sigma}_d$ for all $b$.
    \end{itemize}

    \item \textbf{Sub-covariance extraction}:

    \begin{itemize}
        \item \textbf{Row Selection}: For each combination $(b, d)$, we select the rows corresponding to the indices in $\mathbf{i}_b$:
        \[
        \mathbf{\Sigma}_{\text{rows}}[b, d, :, :] = \mathbf{\Sigma}_{\text{exp}}[b, d, \mathbf{i}_b, :].
        \]

        \item \textbf{Column selection}: We then select the columns corresponding to $\mathbf{i}_b$:
        \[
        \mathbf{\Sigma^k}[b, d, :, :] = \mathbf{\Sigma}_{\text{rows}}[b, d, :, \mathbf{i}_b],
        \]
        resulting in the sub-covariance matrices $\mathbf{\Sigma^k} \in \mathbb{R}^{B \times D \times K \times K}$.
    \end{itemize}
\end{enumerate}

These operations are highly parallelizable, enabling the simultaneous extraction and computation of all required sub-covariance matrices. Moreover, by avoiding explicit loops and leveraging tensor operations, the computational overhead is significantly reduced. This approach scales efficiently with both the number of subsets and the size of the covariance matrices, making it well-suited for large-scale problems.

\subsubsection{Batch processing of multiple orders of interactions using padded batches}
\label{sec:padded_batches}

One limitation of the previously discussed batch calculation of $DTC$, $TC$, $S$, and $\Omega$ is that batched $n$-plets must be of the same order to have the same size. This requirement arises because batched computations can only be broadcast over elements with identical shapes, and the sampled sub-covariance matrix of a given $n$-plet has a shape that depends on the size of the $n$-plet. This issue is common in deep learning models where inputs can have variable lengths, rendering batch processing challenging. A widely recognized foundational approach to handling variable-length inputs in deep learning—especially in the context of language modeling and sequence-to-sequence tasks can be found in early sequence modeling papers \cite{sutskever2014sequence, bahdanau2014neural, vaswani2017attention}. These studies popularized the practice of padding input sequences with special tokens so that all items in a batch share a uniform length, while the model learns to ignore the padding during training. For sub-covariance matrices, we cannot simply add a special token as padding because there are no training mechanisms to learn to ignore them. Instead, we add as padding an identity matrix. Since the added component is independent, we have the following:

\[
\begin{aligned}
H(\Sigma^k)  
&= H(\Sigma^k + I^{n-k}) - H(I^{n-k}) \\
&= H(\Sigma^k + I^{n-k}) -  (n-k)H(N(0,1)) \\
&= H(\Sigma^k + I^{n-k}) -  (n-k) \cdot 1.4189,
\end{aligned}
\]

where $H(\cdot)$ denotes the entropy function, $\Sigma^k$ is the covariance matrix of the $k$ variables of the $n$-plet, 1.4189 is the entropy of a normal distribution with standard deviation 1, and $I^{n-k}$ is the identity matrix of size $n-k$. 

\begin{figure}[h!]
    \centering
    \includegraphics[width=\textwidth]{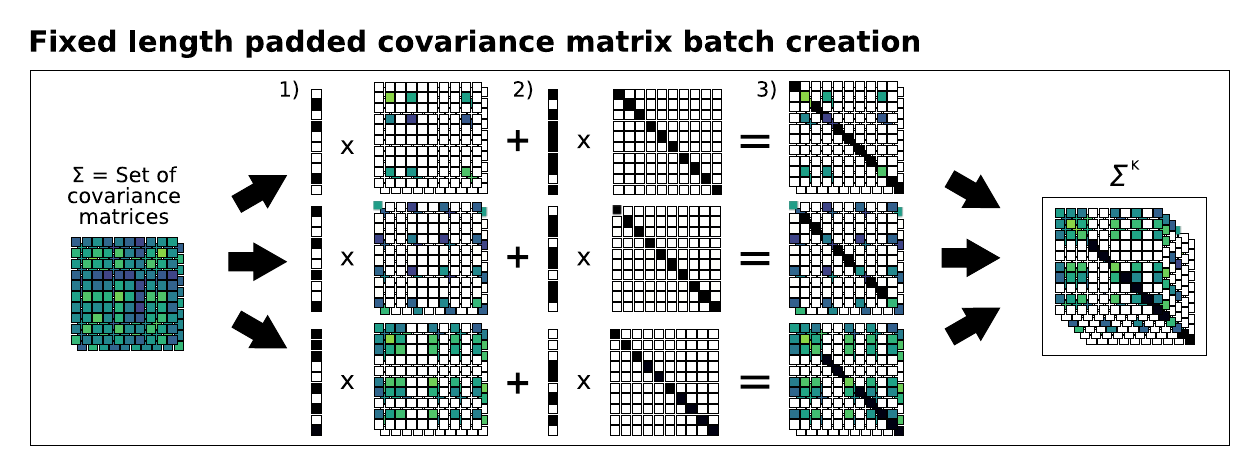}
    \caption{\textbf{Sub-covariance matrices sampled with padding to allow different covariance matrix sizes in a single batch.} 
    1) First, a mask is applied to the full covariance matrices using a masked encoding of the $n$-plets (each with a different number of masked variables) to obtain each sub-covariance matrix. At this point, the obtained covariance matrices are invalid as the masked rows and columns have zeros on the diagonal, yielding a constant distribution. 2) Then, an identity matrix is masked with the inverted $n$-plet encodings. 3) Both masked matrices are added to obtain the final covariance matrix where the rows and columns of the $n$-plet have the values from the full covariance matrix, and the remaining rows and columns have ones on the diagonal and zeros elsewhere, representing an independent standard normal component.}
    \label{fig:padded_batch_process}
\end{figure}

Thus, we can create a batch of padded sub-covariance matrices by adding an independent normally distributed sub-component. We then perform computations in a batched manner as previously explained, and subsequently subtract the entropy of the added component, which is a known constant value multiplied by the length of the padding. Additionally, because the added component corresponds to a standard normal distribution covariance matrix, no bias correction needs to be applied to this sub-component. Figure~\ref{fig:padded_batch_process} illustrates the proposed padding mechanism. It is worth noticing that in order to avoid unnecessary operations, the sampled sub-covariance matrix and the independent components are not sorted to be separated, but maintains the original positions of the variables in the full original covariance matrix. 

While employing a padding strategy enables us to compute $TC$, $DTC$, $\Omega$ and $S$ in a batched fashion, it is more memory-intensive because the batched covariance matrices are larger than the not padded version. Therefore, it is not always advisable to use this strategy. If possible, sorting and processing batches by orders of interaction is preferable to avoid this extra overhead.

\subsection{Heuristics}
\label{sec:heuristics}
Despite pushing the computational limits on the number of variables that can be processed in a reasonable time, systems with a larger number of variables still require strategies to estimate the $\Omega$ as the number of $n$-plets to compute grows exponentially (See ~\ref{sup:combinatorial_explosion}). In this section, we describe two heuristic algorithms implemented in THOI: the greedy algorithm (GA) and simulated annealing (SA). These algorithms also process data in a batched fashion, enhancing their efficiency and allowing us to explore the space of $n$-plets more thoroughly to ensure that the obtained values are representative of the entire set.

\subsubsection{Greedy Algorithm}

GA are a fundamental class of algorithms in computer science and optimization, characterized by constructing a solution iteratively by selecting the best available option at each step based on the current state. They make locally optimal choices with the hope that these choices will lead to a globally optimal solution. Importantly, a GA does not reconsider its previous decisions, meaning it does not backtrack or revise its choices once made.

\paragraph{Key Characteristics}

\begin{itemize}
    \item \textbf{Greedy Choice Property}: At each step, the algorithm makes the most advantageous choice based solely on the current state, without considering future consequences.
    \item \textbf{Optimal Substructure}: The problem can be broken down into smaller sub-problems, and an optimal solution to the overall problem contains optimal solutions to its sub-problems.
\end{itemize}

While GA are powerful for certain types of problems and domains~\cite{cormen2009greedy}, their effectiveness depends on whether the problem exhibits the optimal substructure property. 
%Although this is not the case for all synergistic components—especially in scenarios where hierarchical partitions of synergistic interactions show multiple levels of emergence—in many instances, synergistic components tend to be interconnected, making them suitable for GA.

We designed the GA to efficiently select subsets of variables ($n$-plets) of a given size (order of interaction) that maximize (or minimize) a given function over the $TC$, $DTC$, $\Omega$, and $S$ measures and over a sequence of datasets. Traditional GA often evaluate candidates sequentially; however, we leverage PyTorch's batch processing capabilities to evaluate all possible candidates in parallel at each step. This approach allows the algorithm to systematically expand the current solutions with all possible new candidates, evaluate all the new solutions in a single batched step, and then efficiently select the optimal solutions.

At each iteration, the algorithm expands the current set of $\kappa$ solutions, each of size $t$, by adding every possible candidate variable not already included in the solution. Suppose we have $N$ total variables and each current solution includes $t$ variables; there are $N - t$ candidate variables remaining for each solution. This results in a total of $\kappa \times (N - t)$ new candidate solutions of size $t + 1$.

To manage this expansion efficiently, we represent the solutions and candidates using tensors and perform operations in a batched, matrix-oriented fashion. The process can be described as follows:

\begin{enumerate}
    \item \textbf{Current Solutions Representation}: Let $\mathbf{\chi}$ be a tensor of shape $(\kappa, t)$ representing the current $\kappa$ solutions, where each row corresponds to one solution containing $t$ variable indices.

    \item \textbf{Candidate Variables Identification}: For each current solution $\mathbf{\chi}[j]$, we identify the set of valid candidate variables $\mathbf{K}[j]$ not yet included in $\mathbf{\chi}[j]$. This results in a tensor $\mathbf{K}$ of shape $(\kappa, N - t)$, where $N - t$ is the number of remaining variables.

    \item \textbf{Expansion to New Candidate Solutions}: We create a new tensor $\mathbf{\chi}_{\text{new}}$ of shape $(\kappa, N - t, t + 1)$ to hold all possible expanded solutions. Each element $\mathbf{\chi}_{\text{new}}[j, i]$ is constructed by concatenating the $j$-th current solution $\mathbf{\chi}[j]$ with the $i$-th candidate variable $\mathbf{K}[j, i]$, resulting in a candidate solution of size $t + 1$. This can be all implemented using the same expanding and indexing strategy introduced in section \ref{sec:batched_subcovariance_matrices}.

    \item \textbf{Batch Evaluation}: The tensor $\mathbf{\chi}_{\text{new}}$ contains all possible new candidate solutions formed by adding one variable to each current solution. We can then evaluate all these $\kappa \times (N - t)$ candidate solutions simultaneously using batched operations in PyTorch.

    \item \textbf{Selection of Optimal Solutions}: After evaluating the new candidate solutions, we select the top $\kappa$ solutions based on the optimization criterion (e.g., maximizing or minimizing the $\Omega$ measures). These selected solutions become the current solutions for the next iteration.
\end{enumerate}

 The final GA has the following steps:

\begin{itemize}
    \item \textbf{Solution Initialization}: The algorithm starts with the top $\kappa$ $n$-plets of the initial size, obtained by performing an exhaustive search over that order of interactions.
    \item \textbf{Iterative Incremental Step}: At each iteration, we add one variable to the current $\kappa$ solutions by choosing from the set of variables not yet included, ensuring that the added variables are optimal at each step. This step is implemented using the previously explained strategy of batch expansion, calculation, and evaluation of all possible solutions for all $\kappa$ $n$-plets simultaneously.
    \item \textbf{Termination Criteria}: Once the optimal $n$-plets reach the desired order of interaction, the algorithm returns the obtained solutions.
\end{itemize}

\subsubsection{Simulated Annealing Algorithm}

SA is a stochastic optimization technique inspired by the physical process of annealing in metallurgy, where materials are heated and then slowly cooled to reach a stable state with minimal internal energy~\cite{kirkpatrick1983optimization}. In computational terms, SA explores a solution space by initially allowing for random, high-energy moves to escape local optima, gradually decreasing the probability of such moves over time to settle into a global optimum.

For HOI analysis, SA complements the GA by enabling exploration beyond local optima, thus avoiding the limitations of non-optimal substructures. While a GA focuses on making the best immediate choice at each step, SA introduces randomness, allowing it to occasionally accept suboptimal moves. This flexibility helps the algorithm navigate complex solution landscapes more thoroughly, overcoming the limitations of greedy methods in avoiding local optima and reaching a more globally optimal solution. Furthermore, because SA operates on a solution landscape where each solution is linked with related solutions by certain criteria (e.g., differing by one or more variables), it allows us to explore the space either inside an order of interaction or across order of interactions. To explore the landscape across orders or interactions, we transition from a solution at one order of interaction to a neighboring solution at another order by adding or removing a variable. Our implementation extends the classic SA algorithm by introducing batch processing, enabling the simultaneous optimization of multiple candidate solutions at once. This is achieved by representing the solutions as a batch tensor and provides better explorations of the space.

One challenge in evaluating multiple $n$-plets from different orders of interactions is that batch processing typically assumes all covariance matrices in the batch have the same size. To overcome this, we used the multi order batch processing explained in Section~\ref{sec:padded_batches}, which allows us to evaluate covariance matrices of different sizes in a single batch. The implemented SA algorithm proceeds as follows:

\paragraph{Initialization}

\begin{itemize}
    \item \textbf{Solution Representation}: Each solution is represented as a masked vector of length $N$, where $N$ is system size, a one in the $i$-th position means the $i$-th variable is in the $n$-plet, and a zero means it is not.
    \item \textbf{Initial Solutions}: A batch of $\kappa$ random solutions is generated, with each solution containing at least $3$ elements to be all valid $n$-plets.
\end{itemize}

\paragraph{Energy Evaluation}

\begin{itemize}
    \item \textbf{Objective Function}: An energy (or cost) function $E$ evaluates each solution. This function can be customized based on the problem and supports both maximization and minimization objectives.
    \item \textbf{Batch Evaluation}: The energy of all solutions in the batch is computed simultaneously, exploiting parallelism.
\end{itemize}

\paragraph{Acceptance Criteria}

\begin{enumerate}
    \item \textbf{Energy Difference}: Calculate the change in energy $\Delta E = E_{\text{new}} - E_{\text{current}}$.
    \item \textbf{Acceptance Probability}: A solution is accepted with probability:
    \[
    P = 
    \begin{cases}
    1, & \text{if } \Delta E > 0 \\
    \exp\left(-\dfrac{\Delta E}{Temp}\right), & \text{if } \Delta E \leq 0
    \end{cases}
    \]
    where $Temp$ is the current temperature.
    \item \textbf{Update Rules}: If the new solution is accepted, it replaces the current solution; otherwise, the current solution remains unchanged.
\end{enumerate}

\paragraph{Cooling Schedule}

\begin{itemize}
    \item The temperature $Temp$ is initialized to a high value and decreases at each iteration according to a cooling rate $\alpha$ (e.g., $Temp_{\text{new}} = \alpha Temp_{\text{current}}$ with $0 < \alpha < 1$). Smaller $Temp$ yeld smaller probabilities of accepting solutions with negative energies.
\end{itemize}

\paragraph{Termination Criteria}

\begin{itemize}
    \item The algorithm terminates after a fixed number of iterations or if no improvement is observed over a certain number of iterations (early stopping).
\end{itemize}

%\paragraph{Optimizing over multiple datasets}
Both the GA and the SA approaches optimize an objective function defined over the $TC$, $DTC$, $\Omega$, and $S$ measures using the batch processing mechanism described in Section~\ref{sec:efficient_computation}, they inherently accommodate multiple datasets within a single optimization run. As a result, these methods readily support group-level analyses (e.g. population), facilitating the simultaneous consideration and comparison of multiple datasets in a unified framework.

\subsection{fMRI analysis}
\label{sec:fmr_analysis}
The anesthesia analysis described in Section~\ref{sec:fmr_analysis} was conducted using a publicly available dataset of 17 healthy adults scanned on a 3T Siemens  Ref.~\cite{nemirovsky2023openneuro}. Each participant was recorded during resting-state at four propofol anesthesia levels: Awake (no propofol), Light, Deep, and Recovery (no propofol).
We restricted the analysis to only the Awake and Deep. The dataset was already parcellated into 11 predefined networks, each composed of 5 distinct brain regions, resulting in a total of 55 regions per participant. We did not perform any additional preprocessing steps; all analyses were conducted directly on the data as it was published.
One participant was excluded because their data did not include all 55 regions, leaving a final sample of 16 participants and two conditions (32 datasets in total).

We applied both GA (optimized at each order of interaction) and SA algorithm (optimized across multiple interaction orders) to identify the $n$-plets that either maximize or minimize the paired Cohen’s $d$ effect size between the two conditions.  A Wilcoxon signed rank paired test (each participant underwent the two conditions) with no correction for multiple comparisons was used when comparing between conditions.

\subsection{Complex Systems Dataset analysis}
\label{sec:datasets_analysis}
We employed a publicly available dataset from \cite{cliff2023unifying}, selecting only those datasets containing at most twenty variables (obtaining a total of 920 final datasets). No additional preprocessing steps were performed. For each selected dataset, we exhaustively computed the full suite of HOI measures and subsequently derived the 21 metrics described in section \ref{sec:datasets_analysis}.

\section{Discussion}
\label{sec:discussion}
%\textbf{(subtitles of the discussion to guide the structure; to be removed) }

In this article, we introduced THOI, a novel, accessible and efficient Python library designed to compute HOI in complex systems. By leveraging Gaussian copulas and optimized matrix operations in PyTorch, THOI addresses several key challenges in the estimation of HOI, including the combinatorial explosion and the difficulty of directly estimating probability distributions. Our results demonstrate that THOI significantly outperforms existing tools in terms of computational efficiency and capability to run in ordinary laptops. These findings validate THOI’s utility and open the door to broader applications in the study of complex systems.

THOI’s batch-based architecture integrated with PyTorch enables the parallel processing capabilities of modern CPU, GPU, and TPU architectures. Our comparison of THOI's performance against existing libraries reveals a significant improvement in processing time and memory usage. For example, for a moderate 30-variable system, THOI was able to compute $\Omega$ for all the interactions in a fraction of the time required by alternative open-source libraries, such as those implemented in JIDT, HOI Toolbox and HOI while consuming less than 3Gb of RAM. Therefore, our library offers significant advantages in terms of accessibility, speed and memory efficiency through batch processing. 
It is designed as well to perform population-level and other on-the-fly analyses without storing exponentially scaling results. Furthermore, to provide a more complete description of the informational structure of the data, it computes all four measures ($TC$, $DTC$, $\Omega$ and $S$), while other toolboxes require separate computations for each. This increased efficiency not only makes THOI a practical tool for researchers but also enables a level of scalability that was previously unattainable using traditional approaches.

To validate THOI’s performance, we conducted tests on synthetic datasets with known ground-truth values of $\Omega$, specifically PGMs (see Supplemental Methods \ref{sup:PMGs}). Results shown in Figure \ref{fig:fig6_greedy_and_SA_vs_N} showed that THOI heuristics where able to capture inherent properties of the PGM. When optimizing for redundancy, heuristics led to a linear increase in $\Omega$ because, in a PGM R-system, information is copied across its parts. Therefore, adding more variables of the R-system increases the amount of redundancy linearly. In contrast, optimizing for synergy resulted in an exponential decrease in $\Omega$. This occurred because in a PGM S-system, synergy is established through a collider variable, which if not considered within the sub selection of variables, no synergy would be detected. The independent sub-system and the S sub-system are included almost simultaneously. The close agreement between THOI’s results and the ground-truth values highlights its robustness and reliability as a tool for quantifying HOIs in large complex systems. 

Moreover, our validation using empirical data from fMRI studies further confirmed the practical utility of THOI. When applied to fMRI data from human participants, THOI uncovered qualitative and quantitative changes on brain interdependencies induced by deep anesthesia. These findings align with existing literature, which suggests that HOIs in brain networks may be critical for understanding changes in brain dynamics under different states of consciousness \cite{tononi1994measure, casali2013theoretically}.

Our final demonstration of THOI potential included the exhaustive analysis of all possible interactions on more than 900 datasets in less than 30 minutes in a standard laptop. By analyzing the behavior of key features of the informational structure of HOIs across datasets, we proposed a general framework to characterize complex systems in general. These key features can be easily estimated using the GA and SA and provide complementary information --as revealed by the PCA-- about the multi-order level of interdependence, the multi-order level of independence, the proportion and of synergy-dominated interactions and the overall synergy-redundancy balance.  These results establish THOI as a superior tool in terms of accessibility, speed, and scalability, significantly lowering barriers to better understanding complex interactions in multivariate systems.

%\subsection{Limitations:}
Despite its many advantages, THOI presents several limitations primarily due to its reliance on Gaussian-based methods—the Gaussian estimator of joint entropies and the Gaussian copula approach—which may overlook significant interdependencies present in higher moments or non-Gaussian data distributions. While alternative estimators (e.g. non parametric or kernel based), such as those implemented in JIDT \cite{lizier2014jidt} or the HOI toolbox \cite{neri2024hoi}, could provide more robust HOI estimations, they may compromise performance and scalability, making it challenging to analyze larger systems. Additionally, THOI is not yet optimized for high-performance computing architectures with multiple GPUs, requiring custom scripts to handle this kind of architecture. However, one of THOI's main focus was to be able to run in standard computational setups to ensure accessibility. Future developments should address these Gaussian-related limitations and enhance computational optimizations to better leverage current and emerging HPC technologies, thereby improving the tool’s robustness and scalability without sacrificing accessibility.

%\subsection{Practical Implications and Future Directions:}
In the endeavor to embrace real-world large complex systems, previous studies have already employed heuristic algorithms to optimize $\Omega$. For example, in Ref.~\cite{hourican2024efficient} the authors proposed a SA approach and particle swarm optimization to identify synergy-dominated interactions. THOI introduces a more flexible implementation of SA and a GA capable of searching for optimal $n$-plets for any of the four measures $TC$, $DTC$, $\Omega$, or $S$. This framework also offers flexibility, allowing custom metrics derived from these higher-order interactions (HOI)—such as effect size, classification accuracy, or regression performance—to be optimized across datasets. This adaptability is particularly valuable in scenarios where the measure of HOI that is mathematically optimal may not align with the most effective solution for a specific practical application, as discussed in Ref.~\cite{fernandez2010optimal}.

Despite these advancements, these heuristics remain general-purpose optimizers that do not fully exploit the underlying mathematical properties of the $TC$, $DTC$, $\Omega$, and $S$ measures. Future research could focus on developing more principled, measure-specific optimization strategies. For instance, leveraging the recently introduced gradients of $\Omega$~\cite{scagliarini2023gradients} within a gradient-based optimization framework may significantly refine the search process and yield even more robust results.

The development of THOI has important implications for the study of complex systems across multiple domains. In neuroscience, where HOI are thought to play a crucial role in cognition and consciousness \cite{tononi2016integrated,  herzog2022genuine, herzog2024high, luppi2022synergistic, varley2023multivariate}, THOI offers a powerful tool for uncovering the collective behavior of neural systems, that coupled with mechanistic whole brain models \cite{ herzog2024neural}, could provide novel insights about the biophysical origin of observed high-order interdependencies \cite{gatica2022high}. In macroeconomics, where interconnectedness between financial entities and the non-linear dynamics of the economy are key to understanding crises and systemic risks \cite{battiston2020networks, boccaletti2023structure}, THOI can aid in identifying emergent patterns of behavior that may not be immediately apparent through conventional analysis. Similarly, in the study of the brain \cite{santos2023emergence, santoro2024higher, hindriks2024higher}, ecological networks \cite{golubski2016ecological, grilli2017higher}, social systems  \cite{alvarez2021evolutionary, cencetti2021temporal}, and even musical analysis \cite{medina2021hyperharmonic}. %The ability to quantify HOI offers new insights into the dynamics that shape these and other complex systems. 

% \section{References}
\bibliographystyle{naturemag}
\bibliography{references}

\newpage
\appendix

\section{Supplementary Materials}

%\subsection{Mathematical basis}
\label{sup:matematicals}
\subsection{Key concepts}
\label{sup:key_concepts}
HOI hold the potential to uncover intricate statistical relationships in complex systems by offering a principled way to partition the system information. 
Here, we used and information theoretic approach to HOI, where concepts such as entropy, mutual information, Synergy, and Redundancy are fundamentals.

\paragraph{Entropy}
\label{sup:key_concepts-entropy}
Entropy is a measure of the unpredictability or randomness of a system. In information theory, it represents the average amount of information or surprise produced by a stochastic source of data. Higher entropy indicates more uncertainty or complexity within the system. 

\paragraph{Mutual information}
\label{sup:key_concepts-MI}
Mutual information (MI) is a measure of the amount of information that two random variables share. It quantifies the reduction in uncertainty about one variable given knowledge of another. 

\paragraph{Synergy} 
\label{sup:key_concepts-synergy}
Information that can only be accessed when observing the system as a whole. Occurs when the combined information from multiple variables provides more information than the sum of their individual contributions. 

\paragraph{Redundancy} 
\label{sup:key_concepts-redundancy}
Information that is present in multiple variables. Occurs when copies of the same information can be retrieved from different parts of the system.

\subsection{Entropy analytical expression for multivariate Gaussian variables}
\label{sup:gaussian_entropy_formula}

The entropy formula requires knowledge of the probability density function \( p(X^n) \). In most cases, this function is unknown. However, a notable exception is when the data follows a normal distribution, for which the entropy has a well-established, closed-form analytical expression:

\begin{equation}
\label{eq:gaussian_differential_entropy}
    H(X^n) = \frac{1}{2} \log ((2\pi e)^n |\Sigma|)
\end{equation}

Where $\Sigma$ is the covariance matrix and $|\Sigma|$ its determinant.
% It is derived from injecting the JPDF for a multivariate Gaussian distribution into the differential entropy equation as following:

% \begin{equation}
% \label{gaussian distribution}
%     p(x) = \frac{\exp(-\frac{1}{2} (x-\mu)^T \Sigma^{-1} (x-\mu))}{(2\pi)^{n/2} |\Sigma|^{1/2}} 
% \end{equation}

% where $\mu$ is the mean of the distribution.

% \subsubsection{Estimations from data}
As $H(X^n)$ depends only on the covariance matrix $\Sigma$, the problem of entropy estimation for multivariate Gaussian variables is reduced to the estimation of its covariance matrix. 
However, due to finite sample effects, this estimation is biased. 
Then, for a multivariate system of $n$ variables and $T$ samples, the bias corrector for entropy estimator follows \citep{schurmann2004bias}:

\begin{equation}
\label{eq:bias_corrector}
    \eta(x) = \frac{1}{2} n\; \log \left( \frac{2}{T-1} \right) + \sum^n_j \Psi \left( \frac{T-j}{2} \right)
\end{equation}

where $\Psi$ is the digamma function. This bias is subtracted from the estimated entropy when dealing with experimental data.

\subsection{Estimation of entropy via Gaussian copulas}
\label{sup:gaussian_copula}

Current research has explored the estimation of multivariate mutual information using Gaussian copulas, which provide a flexible and robust approach to capture dependencies between random variables \cite{ma2011mutual, ince2017statistical}. Gaussian copulas allow for the separation of marginal distributions from the joint dependency structure, enabling more accurate entropy estimation, particularly when the underlying distributions deviate from normality \cite{joe2014dependence}.

As delineated in equation \ref{eq:gaussian_differential_entropy}, the Gaussian formula provides a robust framework for systems that adhere to a normal distribution. 
However, real-world data often deviate from this idealized distribution. 
In instances where the distribution of the system is not strictly normal but remains approximately normal, Gaussian copulas offer a powerful alternative 
This approach have been used specially in neuroscience both for bivariate [cite ince, cite] and multivariate problems [cite].

\subsubsection{Copulas}
Formally, a $n$-dimensional copula is the cumulative distribution function (CDF) $C\left(u_{1}, \ldots, u_{n}\right):[0,1]^{n} \rightarrow[0,1]$ of a vector of random variables defined on $[0,1]^{n}$ with uniformly distributed marginals $\mathcal{U}_{[0,1]}$ over [0,1]

\begin{equation}
C\left(u_{1}, \ldots, u_{n}\right)=P\left(U_{1} \leq u_{1}, \ldots, U_{n} \leq u_{n}\right)
\end{equation}
where $U_{i} \sim \mathcal{U}_{[0,1]}$

The Sklar's theorem states that any multivariate distribution can be described by separately specifying the marginal distributions and the copula \cite{sklar1959fonctions}. 
Formally, states that for a $n$-dimensional random vector $X=\left(X_{1}, \ldots, X_{n}\right)$, let $F_{X}$ be its CDF with marginals CDFs $F_i(x) = P(X_i \leq x)$ and a copula function $C:[0,1]^n \rightarrow [0, 1]$, such that $\forall x \in \mathbb{R}^{n}:$

\begin{equation}
F(x_1,\dots, x_l) = C(F_1(x_1),\dots, F_k(x_l))
\end{equation}

If the marginals $F_i$ are continuous then the copula $C$ is unique. 
Therefore, a joint distribution $F_{X}$ can be split into marginals and a copula. Conversely, if $C$ is a copula and $F_{1}, \ldots, F_{n}$ are $\mathrm{CDFs}$, then function $F_{X}=C\left(F_{1}\left(x_{1}\right), \ldots, F_{n}\left(x_{n}\right)\right)$ is a
$n$ dimensional CDF with marginals $F_{1}, \ldots, F_{n}$. 
Sklar's theorem relates the copula to the joint distribution function of the variables $U_{i}=F_{i}\left(X_{i}\right)$.

\begin{equation}
C\left(u_{1}, \ldots, u_{n}\right)=F_{X}\left(F_{1}^{-1}\left(u_{1}\right), \ldots, F_{n}^{-1}\left(u_{n}\right)\right), \quad u_{i} \in[0,1]
\end{equation}
where $F_{i}^{-1}$ are the inverse CDFs. 
For a differentiable copula $C$, we can define the \textit{copula density function} as

\begin{equation}
c\left(u_{1}, \ldots, u_{n}\right)=\frac{\partial^{n}}{\partial u_{1} \ldots \partial u_{n}} C\left(u_{1}, \ldots, u_{n}\right) .
\end{equation}

Consider $u_{i}:=F_{i}\left(x_{i}\right)$ the cumulative distribution functions and $f_{i}$ its corresponding PDFs corresponding to the CDFs $F_{i}(\cdot,$. 
The \textit{copula density} function can be written as

\begin{equation} 
\label{eqcd}
c\left(u_{1}, \ldots, u_{n}\right)=\frac{f_{X}\left(F_{1}^{-1}\left(u_{1}\right), \ldots, F_{d}^{-1}\left(u_{n}\right)\right)}{\prod_{i=1}^{n} f_{i}\left(F_{i}^{-1}\left(u_{i}\right)\right)}
\end{equation}

\subsubsection{Gaussian copulas}

Now let $X \sim \operatorname{N}_{n}(\mathbf{0}, \Sigma),$ where $\Sigma$ is the covariance matrix of $X$. 
Then the corresponding Gaussian copula is defined as
\begin{equation}
C_{P}^{G a u s s}(\mathbf{u}):=\Phi_{\Sigma}\left(\Phi^{-1}\left(u_{1}\right), \ldots, \Phi^{-1}\left(u_{n}\right)\right)
\end{equation}
where $\Phi(\cdot)$ is the standard univariate normal CDF and $\Phi_{\Sigma}(\cdot)$ denotes the joint CDF of $X$.

% It allows for the estimation of a rank-based covariance matrix that captures the underlying dependencies among variables without assuming a normal distribution of the data itself. 

Using a Gaussian copula, we transform the marginal of the dataset to a uniform scale and subsequently apply the inverse of the standard normal cumulative distribution function to achieve normally distributed marginals. 
Then, the covariance matrix of the Gaussian copula transformed can be easily estimated.

Specifically, for data with $T$ samples, the Gaussian copula is can be easily implemented following: 

\begin{equation}
\label{eq:gaussian_copula_covariance_estimator}
    \Sigma_{GC}(X) = \Sigma\left[\Phi\left(\frac{rank(rank(X))}{T+1}\right)\right]
\end{equation}

where $\Sigma$ is the covariance operator that takes a $n-$variate system $X$ with shape $(n, T)$ and computes a covariance matrix of shape $(n, n)$. 
$\Phi$ is the inverse CDF of the standard normal distribution and $rank(\cdot)$ is the ranking function.

Once we obtain this covariance matrix, we can then calculate the Gaussian entropy following Eq. \ref{eq:gaussian_differential_entropy}, which serves as an lower bound for entropy in data that is not normally distributed. This approach leverages the closed form of entropy for Gaussian variables, avoiding costly non parametric estimations of probability density functions by reducing the problem to the estimation of the covariance matrix of the copula-transformed data, which requires much less samples.

%\subsection{Estimation of entropy via Gaussian copulas}
\label{sup:combinatorial_explosion}
%Higher-order interactions (HOI) analysis imposes the aforementioned combinatorial problem, which is the exhaustive computation of all possible combinations of components at all orders of interaction.  This problem poses significant computational challenges, especially in systems characterized by a high number of variables, such as typical complex systems such as the brain, neural networks, or social networks.

\subsection{The combinatorial explosion}
To properly capture the information structure within a system, especially when using O-information, we should consider all possible groupings or "n-plets" of components (i.e. combinations of $n$ variables), from triplets up to the full set of components involved. 
For a system with $n$ variables, the number of possible combinations increases exponentially with $n$. The number of ways to choose $k$ components from $n$ is given by the binomial coefficient $\binom{n}{k}$. From the binomial theorem, we know the following: 

\begin{equation} \label{eq:comb}
\begin{split}
\sum_{k=0}^{n} \binom{n}{k}   & = 2^n  \\
   \binom{n}{0}  + \binom{n}{1}  + \binom{n}{2} + \sum_{k=3}^{n} \binom{n}{k}  & =  2^n   \\
\sum_{k=3}^{n} \binom{n}{k}    & =  2^n -1 -n -\frac{n(n-1)}{2}
\end{split}
\end{equation}

Computing O-information for each combination involves evaluating the joint entropy of variable subsets, requiring significant data manipulation and computational resources. Since joint entropy must be calculated for every possible subset, this exhaustive approach ensures no potential interactions are overlooked, but it imposes a high computational burden. The complexity escalates as the number of variables increases, often making the full computation infeasible. This challenge is especially relevant in fields like neuroscience, where systems may consist of hundreds or thousands of units, such as neurons or brain regions

\subsection{Probabilistic graphical models (PGM)}
\label{sup:PMGs}
To generate data where the ground truth value of the $\Omega$ is known, we use probabilistic graphical models (PGM) as introduced in \cite{koller2009probabilistic}. PGMs are structured representations of the conditional dependencies between random variables in a system. It uses a graph where nodes represent the variables, and edges represent the probabilistic relationships between them. These models have been used to generate synergistic (Head-to-Head) and redundant (Tail-to-Tail) systems, regardless of the  probability distribution \cite{rosas2024characterising}. Here we add a parameter $c$ that controls the level of synergy or redundancy on each system, respectively.

\paragraph{Synergy system (S-system):}
In a S-system configuration, each variable \( X_1, X_2, \dots, X_n \) is marginally independent, and a variable \( Y \) depends on all of them. In this model, arrows in the graphical model point from each \( X_i \) to \( Y \), meaning that \( Y \) is influenced by the joint state of the \( X_i \)'s weighted by the parameter $c$. Mathematically, this model is expressed as:
    \[
    p(X^n, Y) = p(Y \mid X^n, c) \prod_{j=1}^n p(X_j)
    \]

In this case, the variables \( X_1, X_2, \dots, X_n \) are independent, but their collective interaction determines \( Y \), leading to synergistic interaction between any group of variables comprising \( Y \) and at least two different \( X_i\) variables. 

If all variables are Normally distributed, the covariance matrix, $\Sigma_S$, of the S-system follows:

\[
\Sigma_S = 
\begin{bmatrix}
\mathbf{I}_n & c \mathbf{1}_n \\
c \mathbf{1}_n^\top & n c^2 + 1 \\
\end{bmatrix}
\]
where $\mathbf{I}_n$ is the $n \times n$ identity matrix and $\mathbf{1}_n$ is an $n \times 1$ vector of ones.

\paragraph{Redundancy system (R-system):}
In an R-system configuration, variables \( X_1, X_2, \dots, X_n \) are conditionally independent given \( Y \). In this model, arrows in the graphical model point from \( Y \) to each \( X_i \), meaning that \( Y \) acts as a common source weighted by $c$ for all \( X_i \)'s. Mathematically, this model is expressed as:

    \[
    p(X^n, Y) = p(Y) \prod_{j=1}^n p(X_j \mid Y, c)
    \]

This structure suggests that all dependencies among the \( X_i \) variables are explained through \( Y \), leading to redundant information across the variables because they all depend on the same source.

If all variables are Normally distributed, the covariance matrix, $\Sigma_R$, of the R-system follows:

\[
\Sigma_R = 
\begin{bmatrix}
c^2 + 1 & c^2 & \cdots & c^2 & c \\
c^2 & c^2 + 1 & \cdots & c^2 & c \\
\vdots & \vdots & \ddots & \vdots & \vdots \\
c^2 & c^2 & \cdots & c^2 + 1 & c \\
c & c & \cdots & c & 1 \\
\end{bmatrix}
\]

\paragraph{Independent (I):}
In a \textbf{Independent} configuration, variables \( X_1, X_2, \dots, X_n \) are independent variables with some given probabilistic distribution. No statistical interdependence is present.

\paragraph{Concatenated system:}

This system consists in concatenating different systems without interactions between them, being each system either a R-system, a S-system (with some modulation of the $c$ parameter) and a set of independent variables. 
Specifically, we created a concatenated system composed of 5 sub-systems. Two redundant (one weak with $c=0.5$ and one strong with $c=1$), two synergistic (one weak with $c=0.5$ and one strong with $c=1$) and one independent. Each system was composed of 20 variables, creating a system of 100 variables in total.
Because  $\Omega$ satisfies the additive property for independent components ($A \perp B \Rightarrow \Omega(A+B) = \Omega(A) + \Omega(B)$), we can know the $\Omega$ for the whole system. 
Using the aforementioned ground-truth covariance matrices, the entropy equation (eq. \ref{eq:gaussian_differential_entropy}) and THOI, we computes all the ground truth metrics of each system. 

\subsection{Number of repeats in Greedy algorithm}
\begin{figure}[h!] % Positioning options: here, top, bottom, page (htbp)
    \centering
    \includegraphics[width=\textwidth]{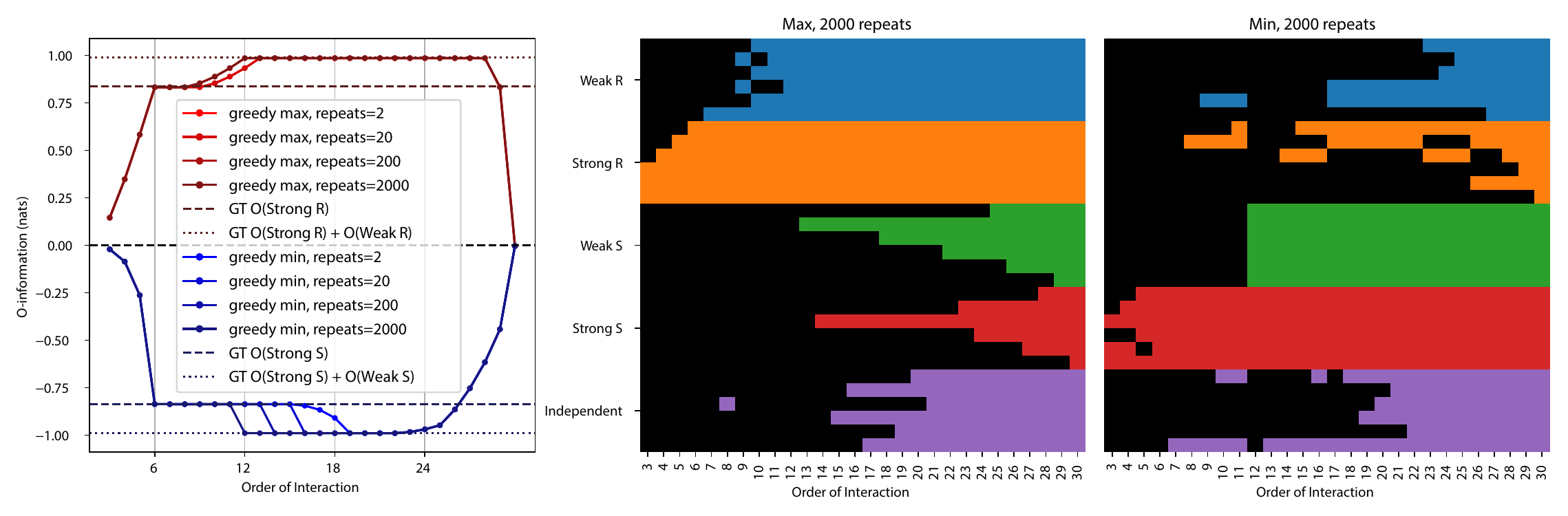 }
    \caption{\textbf{Dependence of the greedy algorithm with the number of repeats}. Same as Figure 3A, C and E, but for a system with N=30 variables. Left panel shows the maximum (red) and minimum (blue) O-information for different number of repeats (initial conditions). Note that for redundancy the solution converges faster than for synergy in terms of the number of repeats. Center and right panel shows the variables involved in the optimal solution for maximization and for minimization, respectively, using 2000 repeats.   
    }
    \label{supp_fig:greedy_vs_repeats_N-30}
\end{figure}

\end{document}